\documentclass[aps,prd,reprint,amsmath,amssymb,showpacs,preprintnumbers,superscriptaddress,nofootinbib,longbibliography]{revtex4-1}

\usepackage{slashed}
\usepackage{bm}
\usepackage{latexsym,amssymb,amsmath,float,url,mathrsfs}
\usepackage{bbold}
\usepackage{latexsym}
\usepackage{graphicx}
\usepackage{epstopdf}
\usepackage{amsmath}
\usepackage{subfigure}
\usepackage{hyperref}
\usepackage{pifont}
\usepackage{multirow}
\usepackage[dvipsnames]{xcolor}
\usepackage{tabularx}

\usepackage{graphicx}
\usepackage{epsfig}
\usepackage{xcolor}
\usepackage{rotating}
\usepackage{amsmath}
\usepackage{braket}
\usepackage{tikz}
\usepackage{pgfplots}
   \pgfplotsset{compat=1.5}
   \usepgfplotslibrary{fillbetween}

\usepackage{feynmp-auto} 

\usepackage{slashed} 
\usepackage[makeroom]{cancel} 

\definecolor{dgreen}{HTML}{008000}

\definecolor{dblue}{HTML}{0000A0}

\newcommand{\X}{\Delta_{dd}}
\newcommand{\Y}{\Delta_{uu}}
\newcommand{\Z}{\Delta_{ud}}
\newcommand{\Sr}{\Phi}

\newcommand\ptwiddle[1]{\mathord{\mathop{#1}\limits^{\scriptscriptstyle(\sim)}}}

\begin{document}
\hfill{ACFI-T18-11}

\title{Electric Dipole Moments from Post-Sphaleron Baryogenesis}

\author{Nicole F. Bell} 
\email{n.bell@unimelb.edu.au}
\affiliation{ARC Centre of Excellence for Particle Physics at the Terascale \\
School of Physics, The University of Melbourne, Victoria 3010, Australia}

\author{Tyler Corbett}
\email{corbett.t.s@gmail.com}
\affiliation{ARC Centre of Excellence for Particle Physics at the Terascale \\
School of Physics, The University of Melbourne, Victoria 3010, Australia}

\author{Michael Nee}
\email{mnee@student.unimelb.edu.au}
\affiliation{ARC Centre of Excellence for Particle Physics at the Terascale \\
School of Physics, The University of Melbourne, Victoria 3010, Australia}

\author{Michael J. Ramsey-Musolf}
\email{mjrm@physics.umass.edu}
\affiliation{Amherst Center for Fundamental Interactions, Department of Physics, University of Massachusetts Amherst, Amherst, MA 01003, USA}

\begin{abstract}
We consider a model in which baryogenesis occurs at low scale, at a temperature below the electroweak phase transition. This model involves new diquark-type scalars which carry baryon number. Baryon number violation is introduced in the scalar potential, permitting $\Delta B=2$ violating process involving Standard Model quarks while avoiding stringent proton decay constraints. Depending on their quantum number assignment, the diquark-type scalars can couple to either right or left handed quarks, or to both. We show that this model can provide a viable explanation of the baryon asymmetry of the universe provided that the coupling to left handed quarks are present. However, the coexistence of couplings to left and right handed quarks introduces important phenomenological constraints on the model, such as radiative contributions to quark masses and the generation of electric dipole moments for nuclei, which probe the CP even and CP odd products of the relevant couplings constants, respectively. We demonstrate that the strongest such constraints arise from electric dipole moment measurements of the neutron and $^{199}$Hg. These constraints are sufficiently strong that, in the absence of an intricate flavor structure, baryogenesis must be dominated by the couplings of the new scalars to left handed quarks.
\end{abstract}

\maketitle
\renewcommand{\baselinestretch}{1.15}
%

\section{Introduction}

A major outstanding question in fundamental physics is the origin of
the cosmological matter-antimatter asymmetry. While the necessary
requirements to generate a baryon asymmetry are the well known
Sakharov conditions, namely (i) violation of baryon number (B), (ii)
violation of C and CP symmetries (CPV) and (iii) departure from
thermal equilibrium, the exact mechanism is yet to be
established. Importantly, any mechanism that adequately fulfills these
three requirements will require the introduction of new particles and
interactions, beyond those that exist in the Standard Model.

A well studied mechanism is that of electroweak
baryogenesis~(EWBG)~\cite{Morrissey:2012db,Trodden:1998ym,Cline:2006ts,Riotto:1998bt} where an asymmetry is generated
via CP violating interactions during a first order electroweak phase
transition with the requisite B violation provided by electroweak
sphalerons.  The SM must be augmented with new physics
to make this scenario viable.  An extended scalar sector is required
to produce a strongly first order phase transition, which is necessary
to fulfill the out-of-equilibrium requirement. In addition, new sources
of CPV must be introduced as the SM Jarlskog factor is suppressed by
small CKM matrix elements and small quark masses.  These challenges notwithstanding,
electroweak baryogenesis has the appealing feature of being a
weak-scale mechanism and thus eminently testable.

This is to be contrasted with high scale baryogenesis mechanisms, such
as the out-of-equilibrium decay of a heavy
particle. Leptogenesis~\cite{Davidson:2008bu,Buchmuller:2005eh,Fukugita:1986hr} (type-I) is a particularly elegant example of such a scenario in which heavy singlet
neutrinos - introduced to allow the seesaw mechanism - decay to lepton
and antilepton final states with unequal rates. The resulting lepton
asymmetry is subsequently converted to a baryon asymmetry via SM
sphaleron processes. While the mechanism is very
simple and requires minimal new physics, the energy scale is
inaccessible to experiments, being $\agt 10^{10}$~GeV in the simplest
scenarios.

In this paper we consider an out-of-equilibrium decay process which
occurs at an experimentally accessible energy scale. In this sense,
the mechanism shares the endearing features of both leptogenesis and
EWBG. Notably, the asymmetry will be generated below the electroweak
phase transition such that sphalerons play no role.  The specific
model we consider is a generalization of the ``post-sphaleron
baryogenesis'' first introduced in Ref.~\cite{Babu:2006xc}, in which the B asymmetry is
directly produced by the decays of a new scalar, $\Phi_r$:
$\Phi_r\rightarrow 6q$ and $\Phi_r\rightarrow 6\overline{q}$.
Importantly, the new physics in this model violates baryon number by
two units, $\Delta B=2$.  This renders the physics safe from the
proton decay constraints that plague $\Delta B=1$ baryon number
violating interactions, such as those in older GUT baryogenesis
models. Because testable low energy baryogenesis mechanisms are relatively rare, this post-sphaleron baryogenesis idea warrants further investigation, which is the purpose of this paper. Similar models have been studied in a non-baryogenesis context in \cite{Arnold:2012sd,Baldes:2011mh,Dev:2015uca,Bowes:1996xy}.

How do we experimentally test this scenario? The existence of $\Delta
B=2$ baryon number violating processes can potentially be observed in
neutron-antineutron oscillation measurements, as was considered in
detail in Ref.~\cite{Babu:2013yca}. A key focus in this paper will
be to probe the CP violating parameters of the model. A well established
low energy technique to probe CP violation is via the electric dipole moments
(EDMs) of atoms, molecules, nucleons and nuclei, as these can be non-zero only in the
presence of CP violation (for recent reviews, see, {\it e.g} Refs.~\cite{Chupp:2017rkp,Engel:2013lsa}) . We shall calculate EDMs together with naturalness constraints on radiative quark mass contributions, which place bounds on CP odd and CP even products, respectively, of the new coupling parameters in our model. These constraints will dictate which of the new couplings can play a role in baryogenesis.

A key difference between our work and that of Refs.~\cite{Babu:2006xc,Babu:2013yca,Babu:2008rq} is the inclusion of couplings to both left handed (LH) and right handed (RH) quarks, rather than RH quarks alone. This has two significant consequences. First, because EDMs require a chirality flip, they will be generated at one loop only when couplings to LH and RH quarks are both present. Second, the size of the baryon asymmetry will depend on whether the interactions involve LH quarks, RH quarks, or both. Like all out-of-equilibrum decay scenarios, baryogenesis requires the interference of a tree level decay amplitude with the absorptive part of a one-loop amplitude. In post-sphaleron bayoogenesis, those loop diagrams involve $W^\pm$ gauge bosons and hence the distinction between couplings to LH or RH quarks is important. Indeed, we shall see that baryogenesis must be dominated by the couplings of the new scalars to LH quarks.

Our work is organized as follows: In Section~\ref{sec:model} we outline the model and the post-sphaleron baryogenesis mechanism. In Section~\ref{sec:edms} we compute constraints on the new coupling constants, using quark mass and nuclear EDM limits, while in  Section~\ref{sec:baryogenesis} we determine the dependence of the CP asymmetries on those coupling parameters and hence discuss the implication of the constraints. We summarize our conclusions in Section~\ref{sec:conclusions}.

\section{The Model}
\label{sec:model}

We introduce three new colored scalar diquarks $\X$,
$\Y$, and $\Z$, which carry baryon number and couple
to quark bilinears. Their quantum numbers and allowed couplings are
outlined in Table~\ref{tab:qnscalars}. In addition, we introduce a
complex scalar field $\Phi$, which is a SM singlet, carries a baryon
number of $B=2$, and has quartic interactions with the colored
scalars of the form:
\begin{equation}
V\supset\frac{\lambda}{2}\Sr \X \Z^2 +\frac{\lambda'}{2}\Sr\Y\X^2\, .
\end{equation}
If the real part of $\Phi$, to which we will refer as $\Phi_r$, obtains a vacuum expectation value $\Delta B=2$ baryon number violating interactions are induced. Further details of the scalar potential have been discussed in \cite{Babu:2006xc,Babu:2013yca} and are relevant to the $N-\bar N$ oscillation calculations therein.

The interactions of SM fermions with the new colored scalars are given by 
\begin{equation}\label{eq:fermL}
\renewcommand\arraystretch{1.5}
\begin{array}{rl}
\mathcal{L}_{\rm Yukawa}=&\frac{h_{\alpha\beta}}{2}\bar K\X(\bar d_R)_{\alpha}(d_R)^c_{\beta} \\
&+\frac{f_{\alpha\beta}}{2}\bar K\Y(\bar u_R)_{\alpha}(u_R)^c_{\beta} \\
&+g_{\alpha\beta}\bar K\Z(\bar u_R)_{\alpha}(d_R)^c_{\beta} \\ 
&+g'_{\alpha\beta}\bar K\Z\epsilon_{ij}(\bar Q_i)_{\alpha}(Q_j)^c_{\beta} 
+h.c.\, ,
\end{array}
\renewcommand\arraystretch{0}
\end{equation}
where $u_R$ and $d_R$ are the usual RH quark fields of
hypercharge $Y=2/3$ and $-1/3$ respectively, while $Q$ is
the LH quark doublet of hypercharge $Y=1/6$.
The $\bar K$ matrices are the generators for $SU(3)_c$ in the sextet representation. Further details of the sextet representation may be found in \cite{Han:2009ya}.
We have used greek characters for flavor indices and the lower case latin characters $i,j$ for indices in the fundamental representation of $SU(2)_L$. We have and will continue to suppress color indices. They can be reinstated as follows:
\begin{equation}
\bar K^A_{ab}\Delta^A\psi_a\chi_b\, ,
\end{equation}
where $A$ corresponds to a sextet color index, and $a,b$ correspond to indices in the fundamental representation. 

Notice that while $\Y$ and $\X$ couple only to RH quarks, $\Z$ couples to both LH and RH quark bilinears.  Although the $\Z QQ$ coupling was not considered in the analyses of \cite{Babu:2006xc,Babu:2013yca}, it cannot be forbidden without imposing additional symmetries\footnote{A UV completion based on a left-right symmetric model eliminates this coupling in \cite{Babu:2006xc,Babu:2013yca}.}. We find that the inclusion of this coupling to LH quarks has an important influence on both the EDM and baryogenesis phenomenology of the model.

\begin{table}[t]
\begin{tabular}{| r | ccc | c |}
\hline
Field&$SU(3)_C$&$SU(2)_L$&$U(1)_Y$& couplings  \\
\hline
$\X$&6&1&-2/3&$d_Rd_R$ \\
$\Y$&6&1&4/3&$u_Ru_R$ \\
$\Z$&6&1&1/3&$u_Rd_R$ \\
$\Sr$&1&1&0&  $\X \Z^2$, $\Y \X^2$ \\
\hline
\end{tabular}
\caption{The quantum number of the new scalars, together with their
  couplings to quark bilinears and the allowed quartic scalar
  interactions.}
\label{tab:qnscalars}
\end{table}

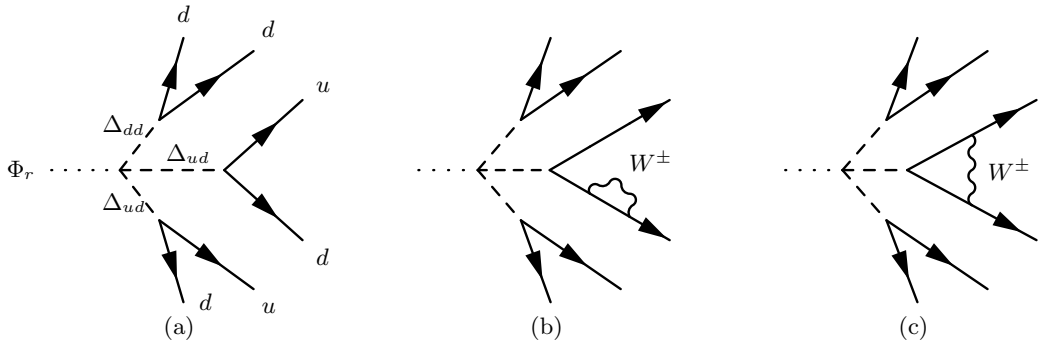
\begin{figure*}
\begin{tabular}{ccccc}
\begin{fmffile}{baryotree}
\begin{fmfgraph*}(100,100)
\fmfsurroundn{v}{12}
\fmf{dots,tension=8}{v7,o1}
\fmf{dashes,tension=3}{o1,o2}
\fmf{dashes,tension=3}{o1,o3}
\fmf{dashes,tension=3}{o1,o4}
\fmf{fermion}{o2,v4}
\fmf{fermion}{o2,v3}
\fmf{fermion}{o3,v11}
\fmf{fermion}{o3,v10}
\fmf{fermion,tension=2}{o4,v12}
\fmf{fermion,tension=2}{o4,v2}
\fmflabel{$\Phi_r$}{v7}
\fmflabel{$u$}{v2}
\fmflabel{$d$}{v12}
\fmflabel{$u$}{v11}
\fmflabel{$d$}{v3}
\fmflabel{$d$}{v4}
\fmfv{l.a=0,label=$d$}{v10}
\fmfv{l.a=-160,label=$\Delta_{dd}$}{o2}
\fmfv{l.a=150,label=$\Delta_{ud}$}{o3}
\fmfv{l.a=155,label=$\Delta_{ud}$}{o4}
\end{fmfgraph*}
\end{fmffile}
&\ \ \ \ \ \ \ \ &
\begin{fmffile}{baryoSE}
\begin{fmfgraph*}(100,100)
\fmfsurroundn{v}{12}
\fmf{dots,tension=8}{v7,o1}
\fmf{dashes,tension=3}{o1,o2}
\fmf{dashes,tension=3}{o1,o3}
\fmf{dashes,tension=3}{o1,o4}
\fmf{fermion}{o2,v4}
\fmf{fermion}{o2,v3}
\fmf{fermion}{o3,v11}
\fmf{fermion}{o3,v10}
\fmf{plain,tension=2.7}{o4,o4a,o4b}
\fmf{fermion,tension=2.7}{o4b,v2}
\fmf{plain,tension=2.7}{o4,o5,o6}
\fmf{fermion,tension=2.7}{o6,v12}
\fmf{photon,left,tension=0,label=$W^\pm$}{o5,o6}
\end{fmfgraph*}
\end{fmffile}
&\ \ \ \ \ \ \ \ &
\begin{fmffile}{baryovertex}
\begin{fmfgraph*}(100,100)
\fmfsurroundn{v}{12}
\fmf{dots,tension=8}{v7,o1}
\fmf{dashes,tension=3}{o1,o2}
\fmf{dashes,tension=3}{o1,o3}
\fmf{dashes,tension=3}{o1,o4}
\fmf{fermion}{o2,v4}
\fmf{fermion}{o2,v3}
\fmf{fermion}{o3,v11}
\fmf{fermion}{o3,v10}
\fmf{fermion,tension=1.5}{o5,v2}
\fmf{plain,tension=1.5}{o5,o4,o6}
\fmf{fermion,tension=1.5}{o6,v12}
\fmf{photon,tension=0,label=$W^\pm$}{o5,o6}
\end{fmfgraph*}
\end{fmffile}
\\
(a)&&(b)&&(c)
\end{tabular}
\caption{Diagrams contributing to the $\Delta B=2$ decay of the real part of the scalar field $\Sr$ into six quarks. The dotted line corresponds to $\Sr_r$, the dashed lines to the new scalars $\X$, $\Y$, and/or $\Z$, the solid lines are the final state quarks, and the one loop diagrams contain an intermediate $W$ boson. Diagram (a) corresponds to the tree level decay, (b) to the one loop self energy correction and (c) to the vertex correction diagram.
Baryogenesis occurs via the interference of the tree diagram with the absorptive part of the loop diagrams. This requires that the $W$ in the loop is put on shell and hence a final state $t$-quark is necessary.
If $\Sr_r$ is replaced with its vacuum expectation value (a) also corresponds to the diagram allowing $n-\bar n$ oscillations at tree level.}\label{fig:baryogenesisdiagrams}
\end{figure*}

Baryogenesis proceeds via the decay of the real part of the $\Sr$ field to six quarks or six anti-quarks, thereby violating baryon number by two units. 
The Feynman diagram for the tree level decay is shown in Figure~\ref{fig:baryogenesisdiagrams} (a). In order to obtain a nonzero baryon asymmetry we require the interference of the tree level decay amplitude with the absorptive part of the one loop decay amplitudes shown in Figure~\ref{fig:baryogenesisdiagrams} (b) and (c). The first corresponds to corrections to the quark wave function via a $W$ exchange while the second corresponds to the vertex correction of the $\Z$ couplings $g$ or $g'$. To obtain the absorptive part required for baryogenesis, we require the $W$ in the loop to be on shell; it is thus necessary to have a final state $t$-quark. We also note that of the three sextet scalars only two are required for successful baryogenesis: either $(\X,\Y)$ or $(\X,\Z)$ are sufficient, the latter being the focus of the studies \cite{Babu:2006xc,Babu:2013yca,Babu:2008rq}. As we are interested in the implications of including the left handed couplings of the $\Z$ boson we will also focus on the $(\X,\Z)$ model throughout this paper.

If the dotted line in Figure~\ref{fig:baryogenesisdiagrams}(a) is replaced with the vacuum expectation value of $\Sr_r$, the resulting diagram describes a mechanism for neutron-antineutron oscillations. This both puts constraints on the relevant couplings of the model as well as motivating future studies of neutron oscillations~\cite{Babu:2013yca}. 

As we shall be interested in low energy observables and in baryogenesis that takes place below the electroweak scale, we choose to work in the quark mass basis. Rotating to the mass basis results in a redefinition of the couplings of Eq.~\ref{eq:fermL}, as detailed in Appendix~\ref{app:flavormass}. The resulting Lagrangian coupling up and down type quarks to $\Z$ is:
\begin{equation}
\mathcal{L}=
\bar K\Z(\bar u'_R)G(d'_R)^c+2\bar K\Z(\bar u'_L)G'(d'_L)^c+\, h.c.\, .
\end{equation}
Capital $G$ and $G'$ are the mass basis analogues of the couplings in Eq.~\ref{eq:fermL}, and the generation indices have been suppressed for simplicity. Although we perform our calculations in the mass basis, we will still reference the properties of the couplings in the flavor basis to derive certain results in Section~\ref{sec:baryogenesis} as the flavor structure is crucial in identifying the leading contributions to the baryon asymmetry. We are now in a position to calculate constraints on the $G$ and $G'$ matrices and the implications of those constraints on the parameter space for successful baryogenesis.

\section{Quark mass and EDM constraints}
\label{sec:edms}

As mentioned above, the introduction of the new LH coupling of the $\Z$ has the potential make significant new contributions to the baryon asymmetry. An important question to address, however, is whether the coexistence of LH and RH couplings will imply new low energy constraints which might render the new contributions to the baryon asymmetry negligibly small. Of particular concern are operators that break chiral symmetry: the Standard Model Yukawa interactions and dipole operators. Both will in general receive new radiative contributions with the simultaneous presence of both LH and RH $\Z$ couplings. In the case of Yukawa interactions, \lq\lq naturalness" considerations imply that radiative corrections to the Yukawa couplings should not be considerably larger in magnitude than their tree-level values. With regard to dipole operators, experimental limits on the EDMs of the neutron and $^{199}$Hg atom imply severe restrictions on the relative phases of the $G$ and $G'$ couplings. In what follows, we analyze both considerations in detail. In doing so, we will work with quantities and the degrees of freedom associated with the Standard Model after electroweak symmetry-breaking: quark masses and quark (chromo-)electric dipole moments\footnote{The resulting constraints are equivalent to those that one would obtain by first considering the SM in its electroweak symmetric phase and then applying the results to the broken phase quantities. }.

\subsection{Radiative Quark Masses and Naturalness}

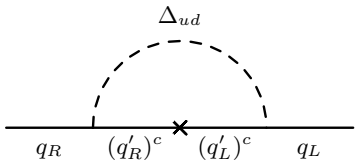
\begin{figure}
\begin{tabular}{cc}
\begin{fmffile}{radmass}
\begin{fmfgraph*}(175,60)
\fmfsurroundn{v}{8}
\fmf{plain,tension=1,label=$q_R$}{v6,o1}
\fmf{plain,tension=1}{o2,o3,o1}
\fmf{plain,tension=1,label=$q_L$}{o2,v8} 
\fmf{dashes,left=1,tension=0,label=$\Z$}{o1,o2}
\fmfv{label=$(q'_R)^c$,l.a=-28}{o1}
\fmfv{label=$(q'_L)^c$,l.a=-152}{o2}
\fmfv{decor.shape=cross,decor.size=7,decoration.angle=0}{o3}
\end{fmfgraph*}
\end{fmffile}
\end{tabular}
\caption{One-loop diagrams contributing to the radiatively generated quark mass. The intermediate quark mass insertion is marked by $\boldsymbol{\times}$ and the dominant contributions come from heavy quarks in the loop.  Only $\Delta_{ud}$, which couples to both LH and RH quarks, can give rise to such mass contributions; there are no analogous diagrams containing $\Delta_{uu}$ or $\Delta_{dd}$.}
\label{fig:radmassdiagram}
\end{figure}

The couplings of the quarks to the scalar $\Z$ give rise to radiative quark mass contributions, as shown in Fig.~\ref{fig:radmassdiagram}. We shall impose a naturalness constraint on the size of these mass contributions, by requiring that the radiatively generated contribution be much smaller in magnitude than the physical quark mass.

Calculating the one loop generated up quark mass at the quark mass scale after running down from the new physics (NP) scale and requiring it be smaller than the tree level quark mass gives the constraint:
\begin{equation}
\delta m_j^u = \frac{4 N_c m_i^d}{8\pi^2}
| G_{ij}^*G_{ij}'|
\log\frac{m_j^u}{M_{\Z}^2}\ll m_j^u\, .
\end{equation}
The constraint for the down quark mass comes from replacing the labels $u\leftrightarrow d$ and changing the order of the indices of $G$ and $G'$. Note there is an implied sum over flavors $i$, but not $j$. We assume that phases in the quark mass matrix, equivalent to a QCD theta term, are removed by assumption of a Pecci-Quinn symmetry. Our constraints thus apply to the magnitude of the radiative mass contributions.

The strongest constraints are associated with intermediate heavy quarks ($i=3$), leading to 
\begin{equation}
\arraycolsep=1.4pt\def\arraystretch{1.5}
\begin{array}{rcl}
| G_{13}^*G_{13}' | &<&10^{-5} \,, \\
| G_{31}^*G_{31}' | &<&10^{-4} \,.
\label{eq:massconstraints}
\end{array}
\end{equation}

Next we consider the EDM constraints which, in contrast with the mass,
are sensitive to the imaginary part of the same products of $G^*$ and $G'$ elements.

\subsection{Electric Dipole Moments Framework}
\label{sec:edmframework}

In studying the constraints from EDM measurements we will closely follow the notation of \cite{Engel:2013lsa}. We quantify the NP contributions to the EDMs in terms of dimension six operators by defining the effective Lagrangian\footnote{Note that we do not explicitly include the dimension-five lepton-number violating operator.},
\begin{equation}\label{eq:efflagrangian}
\mathcal{L}_{\rm eff}=\sum_i \frac{c_i}{\Lambda^2}Q_i
+\mathcal{O} \left( \Lambda^{-3} \right) ,
\end{equation}
where the operators $Q_i$ are constructed from SM fields, and $c_i$ are their Wilson coefficients. The operators admitting CP violation at dimension--six include 10 operators coupling fermions to bosons as well as five four-fermion operators which are enumerated in Tables 2 and 3 of \cite{Engel:2013lsa}. Here we list only those relevant to our analysis,
\begin{eqnarray}
Q_{qG}&=&(\bar Q\sigma^{\mu\nu}T^Aq_R)\ptwiddle{H}G^A_{\mu\nu}\, ,
\label{eq:CPviolatingd6_1}
\\
Q_{qW}&=&(\bar Q\sigma^{\mu\nu}q_R)\tau^i\ptwiddle{H}W_{\mu\nu}^i\, ,
\label{eq:CPviolatingd6_2}
\\
Q_{qB}&=&(\bar Q\sigma^{\mu\nu}q_R)\ptwiddle{H}B_{\mu\nu}\, , 
\label{eq:CPviolatingd6_3}
\\
Q_{quqd}^{(1)}&=&(\bar Q^iu_R)\epsilon_{ij}(\bar Q^jd_R)\, , 
\label{eq:CPviolatingd6_4}
\\
Q_{quqd}^{(8)}&=&(\bar Q^iT^Au_R)\epsilon_{ij}(\bar Q^jT^Ad_R)\, . 
\label{eq:CPviolatingd6_5}
\end{eqnarray}
where $G^{A\mu\nu}$ the gluon field strength tensor with adjoint color index $A$, $W^{i\mu\nu}$ is the $SU(2)$ field strength tensor with adjoint $SU(2)_L$ index $i$, and $B^{\mu\nu}$ is the hypercharge field strength tensor. We denote the SM Higgs doublet as $H$, and define $\tilde H=i\sigma_2H$ where $\sigma_2$ the second Pauli matrix. Here, $q_R$ represents either $u_R$ or $d_R$ and is associated with the ${\tilde H}$ or $H$, respectively. After electroweak symmetry breaking, wherein
\begin{equation}
H^T \rightarrow \left(0,\ v/\sqrt{2}\right)\ \ \ ,
\end{equation}
the operators in Eqs. \ref{eq:CPviolatingd6_2} and \ref{eq:CPviolatingd6_3} generate the dipole couplings to the photon, $Z$, and $W$ bosons. For example, one obtains the effective EDM interaction
\begin{equation}\label{eq:cgamop}
\mathcal{L}_{q\gamma}^\mathrm{EDM}=i\frac{\sqrt{2}v}{2}\frac{\mathrm{Im}[c_{q\gamma}]}{\Lambda^2}\bar q_L\sigma^{\mu\nu} \gamma_5 q F_{\mu\nu}\, ,
\end{equation}
where $F_{\mu\nu}$ is the field strength tensor of the photon field and we have chosen the normalization of the operator coefficient to coincide with that of \cite{Engel:2013lsa}:
\begin{equation}
\mathrm{Im}[c_{q\gamma}] =  \mathrm{Im} [c_{qB}] +  I_3^q\,  \mathrm{Im} [c_{qW}]
\label{eq:imcqgamma}
\end{equation}
with $I_3^q$ being the third component of weak isospin for quark $q$. Note that in the notation of Ref.~\cite{Engel:2013lsa} the coefficients $c_{qB}$ and $c_{qW}$ enter the effective Lagrangian with explicit factors of the U(1)$_Y$ and SU(2)$_L$ gauge couplings, respectively, so that the relation in Eq.~\ref{eq:imcqgamma} carries no explicit dependence on the weak mixing angle.

After integrating out the heavy $\Z$ scalar, we find that the four fermion operators $Q_{quqd}^{(1)}$ and $Q_{quqd}^{(8)}$ are generated at tree-level, while the dipole operators $Q_{fW}$ and $Q_{fB}$ and the chromo-dipole operators $Q_{uG}$ and $Q_{dG}$ arise at one-loop. The four-quark and (chromo-)EDM operators, in turn, contribute to the EDM of the neutron, while only the four-quark and chromo-EDM operators generate potentially significant contributions to the $^{199}$Hg atomic EDM via the nuclear Schiff moment as outlined in \cite{Engel:2013lsa}. 
In the next two subsections, we first derive the relevant operator Wilson coefficients. We then outline the procedure for deriving the implications for the neutron and $^{199}$Hg EDMs and finally apply EDM constraints to the relevant combinations of the $G$ and $G'$  couplings.

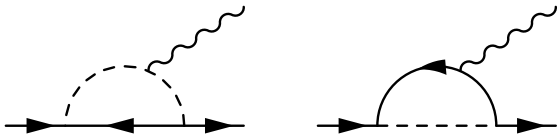
\begin{figure}
\begin{tabular}{c}
\begin{fmffile}{zradiates}
\hspace*{-.5cm}\begin{fmfgraph*}(120,60)
\fmfsurroundn{v}{8}
\fmf{fermion,tension=4}{v6,o1}
\fmf{dashes,tension=1,right=1}{o2,o1}
\fmf{fermion,tension=4}{o2,v8} 
\fmf{fermion,tension=1.0}{o2,o1}
\fmf{phantom,tension=5.0}{v7,o3}
\fmf{phantom,tension=3.0}{o3,v3}
\fmf{photon,tension=2.0}{o3,v2}
\end{fmfgraph*}
\end{fmffile}
\hspace*{-.5cm}\begin{fmffile}{qradiates}
\begin{fmfgraph*}(120,60)
\fmfsurroundn{v}{8}
\fmf{fermion,tension=4}{v6,o1}
\fmf{fermion,tension=1,right=1}{o2,o1}
\fmf{fermion,tension=4}{o2,v8} 
\fmf{dashes,tension=1.0}{o1,o2}
\fmf{phantom,tension=5.0}{v7,o3}
\fmf{phantom,tension=3.0}{o3,v3}
\fmf{photon,tension=2.0}{o3,v2}
\end{fmfgraph*}
\end{fmffile}
\end{tabular}
\caption{One-loop diagrams contributing to the up- and down-quark EDMs. The dashed lines represent the new scalar $\Z$, the external solid lines represent either the up- or down-quark, while the intermediate solid line is summed over all active flavors of quarks. Left: the new heavy scalar $\Z$ radiates a photon or gluon, Right: the intermediate quark radiates a photon or gluon. 
}\label{fig:edmdiagrams}
\end{figure}

\subsection{EDM and chromo-EDM operators at the partonic scale}\label{subsec:treelevelops}
In principle, since both the four-quark and (chromo-) EDM operators arise in our scenario, one must consider contributions from both to the EDMs of diamagnetic systems. In practice, consideration of the four-quark contribution is not presently tractable, due to particularities of the model considered here. In general, the operators $Q_{quqd}^{(1,8)}$ containing only first generation quarks admit an SU(2)$_L\times$SU(2)$_R$ chiral symmetry. Based on this feature, one is able to obtain order-of-magnitude estimates of the hadronic matrix elements relevant to EDMs using na\"ive dimensional analysis (NDA), even in the absence of explicit lattice QCD or model computations. 

As discussed in Appendix B, however, in the basis of weak interaction (flavor) eigenstates, the Wilson coefficients of $Q_{quqd}^{(1,8)}$ vanish for first generation quarks. This result follows from the antisymmetry of the $g'$ couplings with respect to flavor indices. Non-vanishing contributions arise when the flavor of the two RH quarks differs. After rotating to the quark mass basis, one obtains effective four-quark operators involving all first generation quarks as well as those involving other quark flavors. However, operators containing only first generation quarks no longer reflect the SU(2)$_L\times$SU(2)$_R$ chiral symmetry associated with $Q_{quqd}^{(1,8)}$ in the flavor basis, since the operators in the flavor basis also do not reflect this symmetry in our model. Consequently, we are not presently able to estimate the hadronic matrix elements using NDA and considerations of chiral symmetry, as was done in Ref.~\cite{Engel:2013lsa}. Instead, we focus on the loop-induced (chromo-)EDM contributions, using them to obtain conservative upper bounds on the relevant combinations of the mass basis $G$ and $G'$ couplings. Any significant additional contributions from the four quark CPV operators would only strengthen the bounds quoted below in the absence of finely-tuned cancellations among the various conributions.

The quark (chromo-)EDMs and chromo EDMs receive one-loop contributions from the diagrams shown in Fig.~\ref{fig:edmdiagrams} (We note that similar derivation for the $d$-quark EDM alone can be found in \cite{Arnold:2012sd}).
We generate the Feynman rules for our model with the FeynRules mathematica package~\cite{Alloul:2013bka}, calculate the amplitudes in FeynArts~\cite{Hahn:2000kx} and Formcalc~\cite{Hahn:1998yk}, and finally obtain analytic forms for the amplitudes along with their leading order terms in the momentum expansion using Package-X~\cite{Patel:2016fam}.

We obtain
\begin{eqnarray}
{\rm Im}[c_{u\gamma}]&=&-\frac{m_\alpha}{12\pi^2\sqrt{2}v}{\rm Im}[G_{\alpha1}'G_{\alpha1}^*]\left[4+2\ln\left(\frac{m_\alpha^2}{M_{\Z}^2}\right)\right]\, ,\nonumber\\
\label{eq:edm1}\\
{\rm Im}[c_{d\gamma}]&=&\frac{m_\alpha}{6\pi^2\sqrt{2}v}{\rm Im}[G_{1\alpha}'G_{1\alpha}^*]\left[\frac{5}{2}+2\ln\left(\frac{m_\alpha^2}{M_{\Z}^2}\right)\right]\, ,\nonumber\\
\label{eq:edm2}
\end{eqnarray}
where $\alpha$ corresponds to a down-type quark in Eq.~\ref{eq:edm1} and an up-type quark in Eq~\ref{eq:edm2}. We note that for $\alpha=3$ these operators have a strong enhancement from the top and bottom quark masses. We will discuss this feature in more detail below.

The computation of the chromo-EDM is similar, replacing the external photon with a gluon. We find the only difference between the up- and down-quark chromo-EDMs arises from  the difference in mass for each generation and from the couplings $G$ and $G'$.
Using a similar normalization as in Eq.~\ref{eq:efflagrangian} we obtain
\begin{eqnarray}\label{eq:ccedm}
{\rm Im}[c_{uG}]&=&\frac{m_\alpha}{8\pi^2\sqrt{2} v}{\rm Im}[G_{\alpha1}'G_{\alpha1}^*]\ln\left(\frac{m_\alpha^2}{M_{\Z}^2}\right)\, ,\label{eq:cedm1}\\
\nonumber\\
{\rm Im}[c_{dG}]&=&\frac{m_\alpha}{8\pi^2\sqrt{2} v}{\rm Im}[G_{1\alpha}'G_{1\alpha}^*]\ln\left(\frac{m_\alpha^2}{M_{\Z}^2}\right)\, ,\label{eq:cedm2}
\end{eqnarray}

In our analysis below we will make the simplifying assumption that all elements of $G$ and $G'$ are of the same order. Under this assumption we see that the largest contribution to the up and down quark  (chromo) EDMs come from the diagrams with bottom and top quarks in the loops, respectively, due to the dependence on the $m_\alpha$. 
As a result, we expect the limits arising from the $d$-quark EDM to be stronger that those for the $u$-quark EDM by a factor of about ${\rm Im}[c_{dG}]/{\rm Im}[c_{uG}]\sim m_t/m_b\sim40$. This ratio will be observed in the bounds arising from neutron and $^{199}$Hg EDM constraints.

\subsection{EDMs of the neutron and $^{199}$Hg}
\label{sec:analysis}

We now calculate the new contributions to the neutron and $^{199}$Hg EDMs using the operator coefficients determined above. In order to do so we must evolve the coefficients from the heavy scale specified by $M_{\Z}$ to the hadronic scale, which we take to be one GeV.  We obtain the operator coefficients at this scale using the renormalization group analysis of \cite{Dekens:2013zca}, wherein one first evolves down to EW scale integrating out the $t$-quark along the way. Matching is performed while integrating out electroweak gauge bosons and they evolve down to the hadronic scale integrating out the $b$ and $c$ quarks along the way. 

We calculate the neutron EDM, $d_n$, following the procedure of \cite{Engel:2013lsa}:
\begin{equation}\label{eq:dn}
d_n=\sum_{q=u,d}\frac{v^2}{M_{\Z}^2}\left(\beta_{n}^{q\gamma}\, {\rm Im}[c_{q\gamma}]+\beta_{n}^{qG}\, {\rm Im}[c_{uG}]\right) \, ,
\end{equation}
where the ${\rm Im}[c_{q\gamma}]$ are evaluated at the scale $\mu = 1$ GeV and where the hadronic matrix elements are encoded in the $\beta_i^j$. We reproduce their values in Table~\ref{tab:betasgammas} for convenience. Note that the values of the $\beta_n^{q\gamma}$ have been updated to reflect the recent lattice QCD computations of Refs.~\cite{Bhattacharya:2015esa,Bhattacharya:2015wna}. Table~\ref{tab:betasgammas} also gives ranges for the $\beta_i^j$ {\it etc.} to illustrate their degree of uncertainty, however we will use only the central values in our analysis. As pointed out in \cite{Bhattacharya:2015esa,Bhattacharya:2015wna}, one may also anticipate non-negligible contributions from the strange quark EDM, though the magnitude of this contribution is rather uncertain. We have not included this uncertainty in Table~\ref{tab:betasgammas}.

To calculate the EDM of $^{199}$Hg, we begin with the isoscalar and isovector coefficients of the $T$- and $P$-violating pion-nucleon Lagrangian, $\bar g_\pi^{(0)}$ and $\bar g_\pi^{(1)}$ respectively, expressed in terms of our effective operator coefficients:
\begin{equation}
\bar g_\pi^{(i)}=\frac{v^2}{M_{\Z}^2}\gamma_{(i)}^{\pm G}\left({\rm Im}[c_{uG}]\pm{\rm Im}[c_{dG}]\right)   \, 
\end{equation}
where the $\gamma_i$ are given in Table~\ref{tab:betasgammas}. The isotensor contribution, $\bar g_\pi^{(2)}$, is a subleading effect and is neglected in our analysis. The dipole moment of $^{199}$Hg is then given by
\begin{equation}
d_A=\rho_pd_p+\rho_nd_n+ \frac{2m_N g_A \kappa_S}{F_\pi}(a_0\bar g_\pi^{(0)}+a_1\bar g_\pi^{(1)})\, ,
\end{equation}
where $d_p$, the proton EDM, is calculated using Eq.~\ref{eq:dn} with $(\beta^{u\gamma}_n\leftrightarrow\beta^{d\gamma}_n)$ and $(\beta^{uG}_n\leftrightarrow\beta^{dG}_n)$, i.e., by invoking isospin invariance. The last term in the expression above is the contribution from the nuclear Schiff moment. Values assumed for the the nucleon mass $m_N$, the leading order strong interaction coupling $-2g_A/F_\pi$, pion decay constant $F_\pi$, and the coefficients $a_i$ can be found in Table~\ref{tab:parameters}.

\begin{table}
\begin{tabular}{l | c | c}
Coefficient&Best Value $[{\rm e\ fm}]$&Range $[{\rm e\ fm}]$\\
\hline
\hline
$\beta_n^{uG}$&$4\cdot 10^{-4}$&$(1,10)\cdot10^{-4}$\\
$\beta_n^{dG}$&$8\cdot 10^{-4}$&$(2,18)\cdot10^{-4}$\\
\hline
$\beta_n^{u\gamma}$&$1.3\cdot10^{-3}$&$(1.1,1.5)\cdot10^{-3}$\\
$\beta_n^{d\gamma}$&$-4.4\cdot10^{-3}$&$-(4.1,4.7)\cdot10^{-3}$\\
\hline
$\beta_n^{quqd}$&$40\cdot10^{-7}$&$(10,80)\cdot10^{-7}$\\
\hline
$\gamma^{+G}_{(0)}$&$-0.01$&$(-0.03,0.03)$\\
$\gamma^{-G}_{(1)}$&$-0.02$&$-(0.07,0.01)$\\
$\gamma^{quqd}_{(1)}$&$2\cdot10^{-6}$&$(1,10)\cdot10^{-6}$\\
\end{tabular}
\caption{Best values and ranges for the coefficients relevant to the neutron and $^{199}$Hg EDM calculations, all values were taken from~\cite{Engel:2013lsa} except the updated values for $\beta_n^{q\gamma}$ were taken from \cite{Bhattacharya:2015esa,Bhattacharya:2015wna}.}\label{tab:betasgammas}
\end{table}

\begin{table}
\begin{tabular}{r | c | c}
Parameter&Value&Range\\
\hline
\hline
$m_N$&938 MeV&--\\
$g_A$&1.33&--\\
$F_\pi$&185 MeV&--\\
$a_0$&0.01&(0.005,0.05)\\
$a_1$&$\pm$0.02&(-0.03,0.09)\\
$a_2$&0.02&(0.01,0.06)\\
$\rho_p$&$-0.56\cdot10^{-4}$&--\\
$\rho_n$&$-5.3\cdot10^{-4}$&--\\
$\kappa_S$&$-2.8\cdot10^{-4}\ {\rm fm}^{-2}$&--
\end{tabular}
\caption{Parameters values used in the EDM analysis, taken from Tables 4 and 13 of \cite{Engel:2013lsa}. Note that the entry for $\kappa_S$ corrects the overall sign of this quantity from what is given in Ref.~\cite{Engel:2013lsa}.}\label{tab:parameters}
\end{table}

The current 95$\%$ confidence limits on the EDMs of the neutron \cite{Afach:2015sja} and $^{199}$Hg \cite{Graner:2016ses} are
\begin{eqnarray}
\vert d_{\rm n} \vert &\le&3.6\times 10^{-13}\ {\rm e\ fm}\, ,\\
\vert d_{\rm Hg} \vert&\le&7.4\times 10^{-17}\ {\rm e\ fm}\, .
\end{eqnarray}
By demanding that the new physics contributions to these parameter are smaller than the experimental bounds, we will set limits on the $G$ and $G'$ coupling constants. In order to do so, however, we will need to make some simplifying assumptions about the structure of the coupling matrices\footnote{The matrix $G$ has 9 complex parameters, in general, while and $G'$ has the three complex parameters arising from $g'$ plus the four CKM parameters (see the definition of $G'$ in Appendix~\ref{app:flavormass} ). To individually constrain all elements of these matrices would require a significant phenomenological effort, well beyond the scope of this paper. We will therefore make convenient simplifying assumptions about the structure of the matrices. More complex flavor structures are possible, such as those considered in \cite{Babu:2006xc,Babu:2013yca}.}.
We shall assume that all elements of $G$ are of the same order of magnitude, and make a similar assumption about the elements of $G'$. Note, however, that we permit the overall scales of $G$ and $G'$ to  differ significantly.
Given this assumption and that ${\rm Im}[c_{q\gamma,g}]$ are all proportional to $m_\alpha$ (see Eqs.~\ref{eq:edm1} through \ref{eq:cedm2}) we see that the terms enhanced by the heavy quark masses will dominate the limits. Therefore, the EDM constraint directly probe the coupling combinations ${\rm Im}[G_{13}'G_{13}^*]$ and ${\rm Im}[G_{31}'G_{13}^*]$.

\begin{figure}
\begin{tikzpicture}\begin{axis}[
    minor x tick num = 3,
    minor y tick num = 4,
    ymin = -12,ymax = 12,
    xmin = -.5,xmax = .5,
    xlabel={${\rm Im}[G'_{13}G^*_{13}]$},
    xtick = {-.4,-.2,0,.2,.4},
    xticklabels = {$-4\cdot10^{-7}$,$-2\cdot10^{-7}$,$0$,$2\cdot10^{-7}$,$-4\cdot10^{-7}$},
    ylabel={${\rm Im}[G'_{31}G^*_{31}]$},
    ytick = {-10,-5,0,5,10},
    yticklabels = {$-10^{-5}$,$-5\cdot10^{-6}$,$0$,$5\cdot10^{-6}$,$10^{-5}$}
        ]
    
\addplot [blue,domain=-5:5,samples=201,name path=aedmtop]{(2.05 - 23.8*x)};
\addplot [blue,domain=-5:5,samples=201,name path=aedmbottom]{(-2.05 - 23.8*x)};
\addplot [red,domain=-5:5,samples=201,name path=nedmtop]{(20.5 + 35.7*x)};
\addplot [red,domain=-5:5,samples=201,name path=nedmbottom]{(-20.5 + 35.7*x)};

\addplot[red, fill opacity=0.2] fill between [of=nedmtop and nedmbottom,soft clip={domain=-1:1},];

\addplot[blue, fill opacity=0.2] fill between [of=aedmtop and aedmbottom,soft clip={domain=-1:1},];

\end{axis}\end{tikzpicture}
\caption{The blue region is allowed by the $^{199}$Hg EDM constraints, while the red region is allowed by the neutron EDM constraints.  The region allowed by both constraints is shown as purple. We have taken $M_{\Z}=1$ TeV, for larger values of $M_{\Z}$ the results scale roughly as $[{\rm TeV}]^2/M_{\Z}^2$.
}\label{fig:3rdgenonly}
\end{figure}
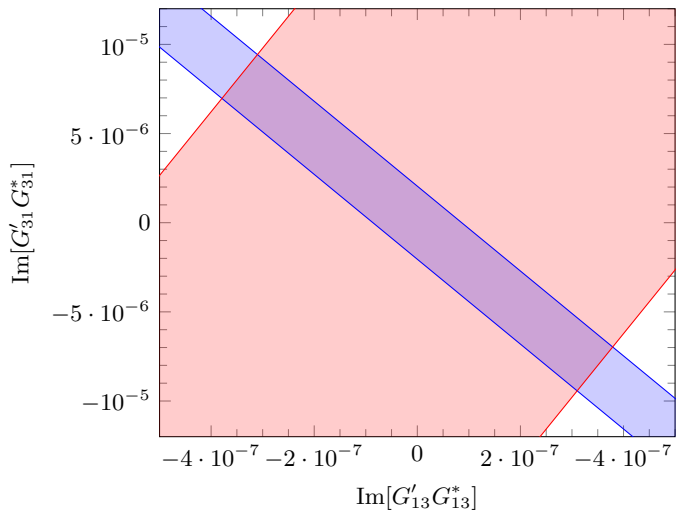

Figure~\ref{fig:3rdgenonly} shows the region in the ${\rm Im}[G_{13}'G_{13}^*]$ vs ${\rm Im}[G_{31}'G_{31}^*]$ plane in which the 95\% confidence level limits of $d_N$ and $d_{\rm Hg}$ are not exceeded. We neglect to perform a more formal likelihood analysis here as we expect our discussion of the baryon asymmetry in Section~\ref{sec:baryogenesis} only to be an order of magnitude estimate. Parameters which pass the neutron EDM constraint are shown in red, while those which pass the $^{199}$Hg constraint are shown in blue. The allowed parameter space is hence given roughly by the region where these bands intersect. We have taken a benchmark value for $M_{\Z}$ of 1 TeV. For larger values of $M_{\Z}$ the results scale approximately as $[{\rm TeV}]^2/M_{\Z}^2$; this naive scaling rule is broken only by the logarithms in the one loop Wilson coefficients of Eqs~\ref{eq:edm1}--\ref{eq:cedm2} and by the running of the effective couplings from the $M_{\Z}$ scale to $M_{\rm QCD}=1$ GeV.

We find that 
\begin{eqnarray}
|{\rm Im}[G_{13}'G_{13}^*]| & < & 4 \times 10^{-7}\,, \nonumber \\
|{\rm Im}[G_{31}'G_{31}^*]|  & < & 1 \times 10^{-5} \,.
\label{eq:EDMconstraints}
\end{eqnarray}
As mentioned above, the difference in the constraints simply reflects the ratio $m_t/m_b\sim40$. For $G_{ij}\sim G'_{ij}$, these bounds correspond to $G$ and $G'$ elements of order $10^{-3}$. 
The constraints coming from the EDMs and those from the radiative quark masses are complementary 
as they constrain the imaginary part and magnitude, respectively, 
of $(G^\dagger m_{u,d}G')_{ii}$. 
From the results presented in Eqs. \ref{eq:massconstraints} and \ref{eq:EDMconstraints}, we conclude that the imaginary parts of these quantities are small, while the real parts can be an order of magnitude or more larger. When we come to the baryogenesis analysis, however, we will simply assume that both the real and imaginary components (and hence the magnitudes) of the $GG'$ quantities satisfy the more stringent of these bounds (Eq.\ref{eq:EDMconstraints}), while the phase is unconstrained, as this 
will lead to the maximum baryon asymmetry\footnote{The  CP-violating asymmetry in $\Phi$ decays is proportional to the imaginary part of a product of couplings, divided by a real part (arising from the tree level decay rate).}.

In the next section we will use these constraints to analyze three main benchmark scenarios for baryon asymmetry generation. These benchmarks are specified by $\Z$ couplings to only RH quarks, only to LH quarks, or to both, namely:
\begin{equation}\label{eq:benchmarks}
\renewcommand\arraystretch{1.5}
\begin{array}{lcl}
G\sim 1&{\rm and}& G'\sim0\, ,\\
G\sim0&{\rm and}& G'\sim1\, ,\\
G\sim 10^{-3}&{\rm and}&G'\sim10^{-3}\, .
\end{array}
\renewcommand\arraystretch{0}
\end{equation} 
The last benchmark corresponds to $GG'\sim10^{-6}$, which is commensurate with the EDM bounds derived above and easily satisfies the radiative mass bounds.

\section{Baryogenesis}
\label{sec:baryogenesis}

\subsection{CP asymmetry of the $\Phi$ decays}
\vspace{-1mm}

The baryon asymmetry is generated at temperatures below the EW phase transition via the out-of-equilibrium decays of $\Sr_r\rightarrow 6q$ and $\Sr_r \rightarrow 6 \overline{q}$, shown in Figure~\ref{fig:baryogenesisdiagrams}. The requisite CP violation is provided by the complex $G$ and $G'$ coupling constants, and through the interference of the tree level decay amplitude with the absorptive part of the 1-loop amplitudes. A key requirement is the presence of $W$ gauge bosons in the loop graphs because, as we shall see below, the non-diagonal flavor structure of the CKM matrix is necessary to generate a non-zero asymmetry. For this reason, there is no analogous $Z$ boson contribution.

As mentioned above, our model differs from that examined in Refs.~\cite{Babu:2006xc,Babu:2013yca,Babu:2008rq}
by the addition of the $G'$ couplings of $\Z$ to LH quarks.  We expect this to allow to larger baryon asymmetries to be generated, because the loop diagrams that permit baryogenesis all involve $W$ bosons. Specifically, if the scalar $\Sr_r$ decays to only RH quarks, then chiral flips are required to permit couplings of the quarks to the $W$, and hence we expect the rates to be suppressed by factors of quark mass. However, if $\Sr_r$ couples to LH quarks, no chiral flip is needed. We therefore expect the baryon asymmetry generated via processes involving RH quarks to be suppressed by factors of $m_{\rm light quark}/M_\Phi$ with respect to the asymmetry generated by processes involving LH quarks.

A full calculation of the baryon asymmetry is beyond the scope of this work. Such a calculation would involve an evaluation of the Boltzmann equations required to track the evolution of the asymmetry, accounting for the $\Sr_r$ decays together with any relevant dilution or washout effects.  We defer such a calculation, together with a detailed exploration of the parameter space, to a future publication.
Instead, as in Refs.~\cite{Babu:2006xc,Babu:2013yca,Babu:2008rq}, we determine the CP asymmetry of the decay rates, defined as
\begin{equation}
\epsilon = \frac{\Gamma({\Sr}\to6q)-\Gamma({\Sr}\to6\bar q)  }{  \Gamma({\Sr}\to6q)+\Gamma({\Sr}\to6\bar q)}\ \ \ .
\end{equation}
The CP asymmetry represents an upper bound on the net baryon asymmetry. In practice, the final baryon asymmetry will be reduced somewhat by dilution or washout effects.

The total $\Sr_r$ decay rate at tree level  is given by:
\begin{equation}\label{eq:treerateggp}
\arraycolsep=1.4pt\def\arraystretch{1.8}
\begin{array}{rcl}
\Gamma_{\rm tree}^\mathrm{TOT}&\equiv&\Gamma({\Sr}\to6q)+\Gamma({\Sr}\to6\bar q)\Big\vert_\mathrm{tree}\\
&=&\frac{1}{\pi^9\cdot 2^{25}\cdot45}12|\lambda|^2{\rm Tr}[H^\dagger H]\frac{M_{\Sr}^{13}}{M_{\Z}^8M_{\X}^4}\boldsymbol{X}^2\, ,\\
\phantom{hi}
\end{array}
\end{equation}
where the phase space factor $1/(45\cdot2^{25}\pi^9)$ has been factored out of the six body phase space \cite{Bashir:2001ad}, and the factor 12 is a color factor. We have defined the quantity $\boldsymbol{X}$, which contains the dependence on the $G$ and $G'$ couplings constants, as 
\begin{eqnarray}
\boldsymbol{X}&\equiv&{\rm Tr}[G^\dagger G+(G')^\dagger G']P_1 \
\nonumber \\
&&
-{\rm Tr}[(G')^\dagger \boldsymbol{m}_u G\boldsymbol{m}_d+G^\dagger \boldsymbol{m}_u G'\boldsymbol{m}_d]\frac{P_0}{M_{\Sr}^2}\, .
\end{eqnarray}
Here and below $\boldsymbol{m}_{u,d}$ are the $3\times3$ diagonal up/down quark mass matrices. 
The quantities $P_1$ and $P_0$ arise from the six body phase space integral. Terms weighted by $P_1$($P_0$) are proportional to phase space integrals of $p_i\cdot p_j$ ($m_i m_j$). Following \cite{Babu:2013yca}  we simplify the momentum dependent integral by assuming the momentum is averaged over the six final state quarks yielding $p_i\cdot p_j\sim M_\Phi^2/6$ and $P_0\sim P_1/6$. We use $P_1 \sim 10^{-4}$, obtained by numerical integration of the 6-body phase space.

There are two dominant contributions to the baryon asymmetry. The first from interference of the tree level decay amplitude with the $t$-quark self energy diagram of Figure~\ref{fig:baryogenesisdiagrams}b, and the second from the interference with the vertex correction of Figure~\ref{fig:baryogenesisdiagrams}c. In both cases a final state $t$-quark is required so that the $W$ boson can be put on shell, which is necessary for obtaining an absorptive part. Evaluating the wavefunction and vertex contributions to the asymmetries, we find
\begin{widetext}
\begin{eqnarray}
\epsilon_{\rm wave}&\sim&
\frac{3g_w^2}{32\pi \boldsymbol{X}M_W^2m_t^2}{\rm Im}\Big([V\boldsymbol{m}_d^2V^\dagger \boldsymbol{m}_u G G^\dagger \boldsymbol{m}_u]_{33}P_1+m_t^2[V\boldsymbol{m}_d^2V^\dagger G' G'^\dagger]_{33}P_1 
\nonumber\\ &&
\phantom{\frac{3g^2}{32\pi \boldsymbol{X}M_W^2m_t^2}{\rm Im}\Bigg([V\boldsymbol{m}_d^2V^\dagger \boldsymbol{m}_u G G^\dagger \boldsymbol{m}_u]_{33}P_1}
-m_t^2[V\boldsymbol{m}_d^2 V^\dagger(\boldsymbol{m}_u G \boldsymbol{m}_d G'^\dagger+G' \boldsymbol{m}_d G^\dagger \boldsymbol{m}_u)]_{33}\frac{P_0}{M_{\Sr}^2}\Bigg)\, , 
\label{eq:sefinal}
\\
\epsilon_{\rm vertex}&\sim&\frac{g_w^2}{32\pi \boldsymbol{X}M_W^2}{\rm Im}
\Bigg([G^*\boldsymbol{m}_dV^\dagger \boldsymbol{m}_u G \boldsymbol{m}_d V^T \boldsymbol{m}_u]_{33}\frac{P_0}{M_\Sr^2}+[G'^*V^\dagger G'\boldsymbol{m}_d^2 V^T]_{33}P_1 
\nonumber\\ &&
\phantom{\frac{g^2}{32\pi \boldsymbol{X}M_W^2}{\rm Im}\Bigg([G^*\boldsymbol{m}_dV^\dagger \boldsymbol{m}_u G \boldsymbol{m}_d V^T \boldsymbol{m}_u]_{33}\frac{P_0}{M_\Sr^2}}
+10\left[G'^* V^\dagger \boldsymbol{m}_u G \boldsymbol{m}_d V^T+G^*\boldsymbol{m}_d V^\dagger G' V^T \boldsymbol{m}_u\right]_{33}P_1\Bigg)\, .
\label{eq:vertfinal}
\end{eqnarray}
\end{widetext}
where $g_{w}$ is the weak coupling constant.
In Eq. \ref{eq:vertfinal}, we have simplifed the expression by replacing logarithms that arise from the momentum integrals with their order of magnitude values, namely, a factor of 1 in the case of the $GG$ or $G'G'$ type terms, and a factor of 10 in the case of the mixed $GG'$ type terms. The full expressions have been used in the numerical analysis below.

\subsection{Dependence of the asymmetry on quark masses}

An important feature of the expressions for $\epsilon_{\rm wave}$ and $\epsilon_{\rm vertex}$ is their dependence on quark masses. Indeed, each of the terms of Eqs.~\ref{eq:sefinal}~and~\ref{eq:vertfinal} have a different quark masses dependence, and this determines which contribution dominates the asymmetry generation. The origin of these mass factors can be traced to either the chiral structure of the quark--$\Z$ couplings, or to the symmetry structure of the $G$ and $G'$ matrices, as we will outline below.

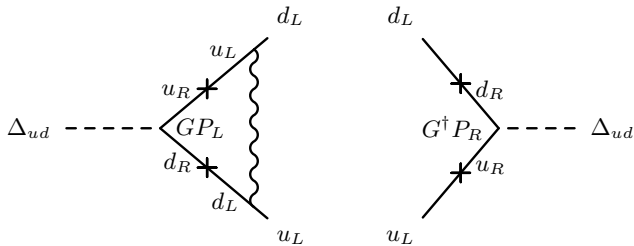
\begin{figure}
\begin{tabular}{cccl}
\begin{fmffile}{vert1vca}
\begin{fmfgraph*}(100,75)
\fmfsurroundn{v}{6}
\fmflabel{$u_L$}{v6}
\fmflabel{$d_L$}{v2}
\fmflabel{$\Z$}{v4}
\fmf{dashes,tension=1}{v4,o1}
\fmf{plain,tension=1}{o1,i11,i12}
\fmf{plain,tension=1}{o1,i21,i22}
\fmf{plain,tension=3.5}{i12,v2}
\fmf{plain,tension=3.5}{i22,v6}
\fmf{photon,tension=0}{i12,i22}
\fmfv{label=$GP_L$,l.a=0}{o1}
\fmfv{decor.shape=cross,decor.size=7,decoration.angle=45}{i11}
\fmfv{decor.shape=cross,decor.size=7,decoration.angle=45}{i21}
\fmfv{label=$u_R$,l.a=190}{i11}
\fmfv{label=$d_R$,l.a=160}{i21}
\fmfv{label=$u_L$,l.a=190}{i12}
\fmfv{label=$d_L$,l.a=160}{i22}
\end{fmfgraph*}
\end{fmffile}
&&
\begin{fmffile}{vert1ccaa}
\begin{fmfgraph*}(75,75)
\fmfsurroundn{v}{6}
\fmf{dashes,tension=1}{v1,o1}
\fmf{plain,tension=1}{v3,o2}
\fmf{plain,tension=1}{o2,o1,o3}
\fmf{plain,tension=1}{o3,v5}
\fmfv{decor.shape=cross,decor.size=7,decoration.angle=45}{o2}
\fmfv{decor.shape=cross,decor.size=7,decoration.angle=45}{o3}
\fmflabel{$d_L$}{v3}
\fmflabel{$u_L$}{v5}
\fmflabel{$\Z$}{v1}
\fmfv{label=$d_R$,l.a=-22}{o2}
\fmfv{label=$u_R$,l.a=20}{o3}
\fmfv{label=$G^\dagger P_R$,l.a=180}{o1}
\end{fmfgraph*}
\end{fmffile}
\end{tabular}
\caption{A diagrammatic explanation of the mass dependence of the vertex correction, in the case where the $\Z$ couples only to RH quarks.
The necessity for four mass insertions can be understood as follows:  $G$ couples a $\Z$ to two right-handed quarks, hence two mass insertions are required so that the quarks can exchange a $W$-boson. Then, in order for the helicities to correctly match the conjugated tree-level diagram, two more mass insertions are required. 
The corresponding expression for the asymmetry $\epsilon_{\rm vertex}$ is found to differ from that \cite{Babu:2006xc} and \cite{Babu:2013yca}, but to agree with that of \cite{Babu:2008rq}.}\label{fig:helicityvertexcorrection}
\end{figure}

The way in which the chiral nature of the couplings is related to the mass factors of $\epsilon_{\rm wave}$ and $\epsilon_{\rm vertex}$ can be easily understood. If the $G$ coupling to RH quarks is involved, chiral flips are required in order to couple the quarks to the $W$ in the loop diagrams. No such chiral flip is needed in the case of the $G'$ coupling to LH quarks.
Therefore, the term proportional to $GG$ requires four mass insertions for the vertex correction\footnote{The necessity of four mass insertions appears to disagree with the expressions of \cite{Babu:2006xc,Babu:2013yca}, but does agree with the work of the same authors in \cite{Babu:2008rq}.}, as illustrated in Fig.~\ref{fig:helicityvertexcorrection}, and two for the wave function correction. The $GG'$ cross term requires only two mass insertions for both the vertex and wave function corrections, while the $G'G'$ term requires no chiral flips for either the wave or vertex topologies.

Inspecting Eq.~\ref{eq:sefinal} we see that the wavefunction corrections appear to require two additional mass quark factors beyond those stipulated above. 
This occurs because, in the limit that the (down-type) quark in the wave function loop diagram is taken to be massless, the amplitudes all contain a factor of $(V^\dagger V)_{ij}=\delta_{ij}$. This renders the remaining product of coupling factors real.
However, by retaining the next order term in a small mass expansion we obtain a 
$(V^\dagger m_d^2 V)_{ij} \ne \delta_{ij}$ dependence, and hence a product of couplings that, in general, will be complex\footnote{More generally, it is clear that distinct quarks masses and non trivial $V_{\rm CKM}$ mixing are important requirements for successful baryogenesis in this model.}. 

For a similar reason, the $G'G'$ vertex contribution also has a factor of $m_d^2$.
Again, in the limit of zero quark mass, we would have 
\begin{equation}
(G'^* V^\dagger G' V^T)_{33}=-\left|V_{\rm CKM}^*[g'^*-g'^\dagger]\right|_{3i}^2={\rm Real}\, ,
\end{equation}
due to the antisymmetry of $g'$. (Note also that $V_{\rm CKM} \rightarrow \mathbb{1}$ in the zero quark mass limit.) Hence we are again required to keep the next order term in quark mass (arising from the cut loop integral) which yields a factor of $m_d^2$ and results in a coefficient which, in general, is complex. 

The quark mass factors in the various contributions to $\epsilon_{\rm wave}$ and $\epsilon_{\rm vertex}$ will determine which terms dominate the generation of the baryon asymmetry.  We will now examine this in the context of the 3 benchmark coupling scenarios which satisfy our constraints from radiative quark masses and EDMs.

%
%

\subsection{Size of the asymmetry for benchmark scenarios}

\begin{table*}
\begin{tabular}{| r || c | c || c | c |}
\hline
&$\epsilon_{\rm wave}(M_{\Sr}=200\ {\rm GeV})$
&$\epsilon_{\rm vertex}(M_{\Sr}=200\ {\rm GeV})$\\
\hline
\hline
$G_{\alpha\beta}\sim1$, $G'_{\alpha\beta}=0$&$10^{-9}$&$10^{-11}$\\
$G'_{\alpha\beta}\sim1$, $G_{\alpha\beta}=0$&$10^{-7}$&$10^{-6}$\\
$G_{\alpha\beta}\sim G'_{\alpha\beta}\sim10^{-3}$&$10^{-7}$&$10^{-6}$\\
\hline
\end{tabular}
\caption{Typical sizes of the CP asymmetry in the three benchmark scenarios. The first corresponds to the assumptions of \cite{Babu:2013yca,Babu:2006xc} in which the $\Z$ couples only to RH quarks, while the second corresponds to the scenario in which  the $\Z$ couples only to LH quarks. In the final scenario, the $\Z$ couples to both RH and LH quarks with similar strength.
In the case of $G'=0$ the asymmetries are smaller than those of \cite{Babu:2006xc,Babu:2008rq}, which adopt a different flavor structure for $G$ and is missing quark mass factors.}
\label{tab:asymm}
\end{table*}

To determine the size of the baryon asymmetry in the presence of both the LH and RH couplings, we take the three benchmark scenarios defined by Eq. \ref{eq:benchmarks}. We generate random complex matrices which satisfy the appropriate symmetries\footnote{See Appendix~\ref{app:flavormass} for a discussion of the symmetries of $g'$.} of $G$ and $g'$, where all nonzero elements are of the order of magnitude specified by Eq.\ref{eq:benchmarks}. We then transform $g'$ to $G'$ by rotating to the mass basis, and compute the values $\epsilon_{\rm wave}$ and $\epsilon_{\rm vertex}$ according to  Eqs.~\ref{eq:sefinal}-\ref{eq:vertfinal}.  The typical order of magnitude obtained via this procedure is recorded in Table~\ref{tab:asymm} for each contribution to the total asymmetry.

In the case of the wavefunction contribution, the term proportional to $G'G'$ (LH couplings) is weighted by $m_t^2$ while the term proportion to $GG$ (RH couplings) is weighted by $m_u m_t$.  In the latter, $m_u$ may be either a $u$-quark or $c$-quark, as $m_u=m_t$ would lead to an overall real value.  We therefore expect that the $GG$ contribution to be suppressed with respect to the $G'G'$ contribution by a factor of $m_c/m_t \sim 10^{-2}$.

For the case of the vertex contribution, we instead expect that the $GG$ term is suppressed with respect to the $G'G'$ term by a factor of $(m_c m_t V^\dagger_{32})/(6 M_\Sr^2 V^\dagger_{21}) \sim 10^{-4}$.  By comparing the first two lines of Table~\ref{tab:asymm} we see that these relative suppression factors are indeed reflected in the values of $\epsilon$ obtained. 
When $G \sim G'$, the $\epsilon$ are dominated by the $GG'$ cross terms and the $G'G'$ contributions.

An important feature of the $\epsilon$'s is that they are insensitive to the overall magnitude of the couplings $G$ and $G'$.  This is because both the numerator and denominator (via the terms denoted $\boldsymbol{X}$) carry two powers of $G$ or $G'$.  However, it is not viable to make the couplings arbitrarily small because this would suppress the magnitude of the tree level decay rate, resulting in a large dilution factor and thus a final baryon asymmetry that is too small. Hence while the $G \sim G' \sim 10^{-3}$ benchmark (or, indeed, even smaller values of $G$ and $G'$) would appear to permit a CP-violating asymmetry, $\epsilon$, of sufficient size, it may be difficult to obtain a suitable tree level decay rate when all couplings are very small.

From our simple analysis, we are able to conclude that the $G=0$, $G'\sim 1$,  benchmark, in which the $\Z$ couples only to LH quarks, provides the most viable baryogenesis scenario and permits CP-violating asymmetries up to $\epsilon \sim 10^{-6}$ before any washout or dilution effects are incorporated.  The $G\sim 1$, $G=0$ benchmark, in which the $\Z$ couples only to RH quarks, is less viable because the CP-violating asymmetries are suppressed by additional factors of light quark mass and/or small CKM matrix elements.  
The $G \sim G'$ benchmark, which features comparable couplings to LH and RH quarks, is disfavoured because the requirement that the magnitudes of the couplings the satisfy the EDM constraints yields too small a tree-level decay rate.

It is important to note that our analysis assumes the coupling matrices have no strong flavor hierarchies. While the adoption of a specific flavor structure could relax our conclusions, a detailed exploration is beyond the scope of this paper. We also note that the CPV asymmetry in this scenario appears at second order in the B-violating coupling.  Under certain assumptions, the requirements of CPT invariance and unitarity -- expressed via the so-called Nanopoulos-Weinberg theorem~\cite{Nanopoulos:1979gx,Kolb:1979qa} -- imply that the asymmetry should vanish at this order. Recently, however, it has been shown in general that this theorem may be evaded in the presence of additional interactions and decay modes~\cite{Bhattacharya:2011sy}, thereby allowing for a non-vanishing asymmetry at second order in the B-violating coupling. A specific application to a simplified version of the model considered here is given in the Appendix A of Ref~\cite{Babu:2013yca}. We defer a more detailed analysis of this issue to our future work.

\section{Conclusions}
\label{sec:conclusions}

In this paper we have examined the ``Post Sphaleron Baryogenesis'' mechanism first introduced in \cite{Babu:2006xc}. While the original scenario featured the coupling of scalar diquarks to RH quarks, we extended the model by the addition of couplings to left-handed quarks, $\Z\bar QQ^c$, which results in a greatly enhanced baryon asymmetry. This occurs simply because baryogenesis proceeds via $W$ loop diagrams, for which contributions involving LH quarks in general require less quark mass insertions than those involving RH quarks.

This comes at the cost, at least in the scenario in which both LH and RH couplings are present, of induced electric dipole moments of nucleons and atoms and radiative contributions to the quark masses. We find that the electric dipole moments of the neutron and $^{199}$Hg provide the most stringent constraints. Under the assumption that there is no significant hierarchy among the elements of LH coupling constant matrix $G'$ (and likewise the RH couplings $G$) 
and that the real and imaginary components of these couplings are comparable, the EDM constraints imply that the $GG'$ couplings constant products are in the $<10^{-5}-10^{-7}$ range. Hence either $G$, $G'$, or both must be small. 

We derived expressions for the self energy and vertex correction contributions to the baryon asymmetry generation, in the presence of couplings to both LH and RH quarks. This allowed up to estimate the relative size of the baryon asymmetry in three benchmark scenarios for which either the $G$, $G'$, or both couplings are non-zero. The most viable scenario was found to be that where only the $G'$ couplings to LH quarks are present. The scenario with only $G$ couplings to RH quarks is disfavoured because additional factors of light quark mass suppress the size of the baryon asymmetry generated, while the scenario in which $G$ and $G'$ are both nonzero is disfavoured by the EDM constraints. Therefore, our new couplings of the diquark scalars to LH quarks is critical for the successful generation of a baryon asymmetry in this model.


\section*{Acknowledgments}
NFB, TC, and MN were supported in part by the Australian Research Council. MJRM was supported in part under U.S. Department of Energy contract DE-SC0011095. This work was performed in part at the Aspen Center for Physics, which is supported by National Science Foundation grant PHY-1607611.

\appendix

\section{Flavor Symmetries and the Mass Basis}\label{app:flavormass}

We note that the coefficients $h$, $f$, $g$, and $g'$ are in general complex $3\times3$ matrices. Consequently, the couplings of the $\Delta$ fields to SM fermion bilinears allow for the possibility of CP-violation. 
We may infer further symmetries of these coupling matrices by recalling the sextet representation of $SU(3)$ is symmetric and applying the identity
\begin{equation}
\bar\psi\chi^c=\bar\chi\psi^c\, .
\end{equation}
From here we see that the coupling $h$ ($f$) which couples a sextet scalar to $\bar d_Rd_R^c$ ($\bar u_Ru_R^c$) is symmetric in flavor. Expanding the LH coupling of the $\Z$ in its $SU(2)_L$ indices and using the above identity as well as that the sextet is symmetric in color allows us to conclude that the matrix $g'$ is antisymmetric. There are no such symmetries of the RH coupling matrix $g$. As we will see below rotating to the mass basis obfuscates some of these symmetries.

A symmetric $3\times3$ complex matrix contains 6 complex couplings while an antisymmetric $3\times3$ complex matrix has three. Thus, in total we have $6+6+3+9=24$ complex couplings ($f+h+g'+g$) in our model. These are not angles and phases as in the case of the CKM or PMNS matrices as these matrices are not unitary. They are instead a priori unconstrained magnitudes and phases. As there is no remaining freedom to remove further phases by field redefinitions of the quarks we are unable to reduce the size of this set.

As the quantities we calculate in the main text are relevant below EWSB we must rotate to the mass basis. Therefore we rotate the Lagrangian of Eq.~\ref{eq:fermL} to the mass basis. We begin in a flavor diagonal basis where the $d$-quark mass matrix is also diagonal. Then we diagonalize the $u$-quark mass matrix by making the following rotations,
\begin{equation}
\renewcommand\arraystretch{1.5}
\begin{array}{rcl}
u_{R,\alpha}&\to& (T_u)_{\alpha\beta}u_{R,\beta}'\, ,\\
u_{L,\alpha}&\to& (S_u)_{\alpha\beta}u_{L,\beta}'=(V_{\rm CKM})_{\alpha\beta}u_{L,\beta}'\, ,
\end{array}
\renewcommand\arraystretch{0}
\end{equation}
where the primed fields are the fields in the mass basis, and we have noted that starting from a diagonal $d$-quark mass matrix allows us to identify the rotation matrix $S_u$ with the CKM matrix. Performing this rotation on the Lagrangian of Eq.~\ref{eq:fermL} we find,
\begin{equation}
\renewcommand\arraystretch{1.8}
\begin{array}{rl}
\mathcal{L}\equiv&\frac{\bar K}{2}\X(\bar d'_R) H(d'_R)^c+\frac{\bar K}{2}\Y(\bar u'_R)F(u'_R)^c\\
&+\, \bar K\Z(\bar u'_R)G(d'_R)^c+2\bar K\Z(\bar u'_L)G'(d'_L)^c\\
&+\, h.c.\, .
\end{array}
\renewcommand\arraystretch{0}
\end{equation}
Here we have suppressed the flavor indices and introduced the matrices $H$, $F$, $G$, and $G'$, which are the analogs of the flavor matrices in the mass basis and are given by:
\begin{equation}
\renewcommand\arraystretch{1.5}
\begin{array}{rcl}
H_{\alpha\beta}&\equiv&h_{\alpha\beta}\, \\
F_{\alpha\beta}&\equiv&(T_u^\dagger)_{\alpha\sigma}f_{\sigma\rho}(T_u^\dagger)_{\rho\beta}\, \\
G_{\alpha\beta}&\equiv&(T_u^\dagger)_{\alpha\sigma}g_{\sigma\beta}\, \\
G'_{\alpha\beta}&\equiv&(V_{\rm CKM})_{\alpha\sigma}[g'_{\sigma\beta}-(g')^T_{\sigma\beta}]\, .
\end{array}
\renewcommand\arraystretch{0}
\end{equation}
We note that the asymmetry manifest in the matrix $g'$ is no longer present in the matrix $G'$. Having rotated to the mass basis we are now free to make calculations in perturbation theory.

\section{EDMs from tree level effective operators}\label{sec:treelevelEFT}

We proceed to integrate out the scalar $\Z$ at tree level by following the procedure outlined in \cite{Henning:2014wua}. This procedure is done above EWSB so that we can first match on to the effective operators contained in the review \cite{Engel:2013lsa}.
Assuming that operators of dimension greater than six have negligible effects on the low energy physics allows us to neglect terms of operator mass dimension seven and higher in the expansion. 
The effective Lagrangian to dimension-six at tree level is given by:
\begin{widetext}
\begin{eqnarray}
\label{eqapp:d6tree}
\mathcal{L}_{\rm Eff}'= && -\frac{g_{\alpha\rho}g_{\sigma\beta}^*}{M_{\Z}^2}
 \left[ \frac{2}{3} \left(Q_{ud}^{(1)}\right)_{\alpha\beta\rho\sigma}+\left(Q_{ud}^{(8)}\right)_{\alpha\beta\rho\sigma} \right]
+\frac{[g'_{\alpha\rho}-g'_{\rho\alpha}][(g')^*_{\sigma\beta}-(g')^*_{\beta\sigma}]}{4M_{\Z}^2}
 \left[ \left(Q_{qq}^{(1)}\right)_{\alpha\beta\rho\sigma} + \left(Q_{qq}^{(8)}\right)_{\alpha\beta\rho\sigma} \right]
\nonumber \\
&& + \frac{g_{\alpha\beta}[(g')^*_{\sigma\rho}-(g')^*_{\rho\sigma}]}{3M_{\Z}^2}
 \left[ 2 \left(Q_{quqd}^{(1)}\right)^\dagger_{\alpha\rho\beta\sigma}
+  \left(Q_{quqd}^{(8)}\right)^\dagger_{\alpha\rho\beta\sigma}+h.c. \right]\, .
\end{eqnarray}
\end{widetext}
Where the operators $Q$ are defined as:
\begin{eqnarray}
\left(Q_{ud}^{(1)}\right)_{\alpha\beta\rho\sigma}&=&(\bar u_\alpha\gamma_\mu u_\beta)(\bar d_\rho\gamma^\mu d_\sigma)\nonumber\\
\left(Q_{ud}^{(8)}\right)_{\alpha\beta\rho\sigma}&=&(\bar u_\alpha\gamma_\mu T^A u_\beta)(\bar d_\rho\gamma^\mu T^A d_\sigma)\\
\left(Q_{qq}^{(1)}\right)_{\alpha\beta\rho\sigma}&=&(\bar q_\alpha\gamma_\mu q_\beta)(\bar q_\rho \gamma^\mu q_\sigma)\nonumber\\
\left(Q_{qq}^{(8)}\right)_{\alpha\beta\rho\sigma}&=&(\bar q_\alpha\gamma_\mu T^Aq_\beta)(\bar q_\rho \gamma^\mu T^A q_\sigma)\nonumber
\end{eqnarray}
And $Q_{quqd}^{(1,8)}$ were defined in Eqs.~\ref{eq:CPviolatingd6_4}~and~\ref{eq:CPviolatingd6_5}. We note that a linear combination of the $CP$-violating $Q_{quqd}^{(1,8)}$ is obtained as a result of our application of the following Fierz identities:
\begin{equation}\label{eqapp:fierz}
\arraycolsep=1.4pt\def\arraystretch{1.8}
\begin{array}{rcl}
\epsilon_{jm}\epsilon_{kn}&=&-\tau_{jk}^J\tau_{mn}^J+\delta_{jn}\delta_{mk}\\
K_{ab}^A\bar K_A^{cd}&=&\frac{1}{2}(\delta_{ad}\delta_{bc}+\delta_{ac}\delta_{bd})\\
\lambda^I_{ab}\lambda^I_{cd}&=&2\delta_{ad}\delta_{bc}-\frac{2}{3}\delta_{ab}\delta_{cd}\\
(\bar\psi_1\sigma^{\mu\nu}P_L\psi_2)(\bar\psi_3\sigma_{\mu\nu}P_L\psi_4)&=&-8(\bar\psi_1P_L\psi_4)(\bar\psi_3P_L\psi_2)\\
&&-4(\bar\psi_1P_L\psi_2)(\bar\psi_3P_L\psi_4)\, .
\end{array}
\end{equation}
It is interesting to note that the asymmetry of the matrix $g'$ is manifest in the operator coefficients. 
Moreover, as only the purely first generation operator contributes to the neutron and $^{199}$Hg EDMs we might na\"ively think the coefficient vanishes due to this antisymmetry in the flavor indices. However, Eq. \ref{eqapp:d6tree} is in the flavor basis and we must now rotate to the mass basis drawing any conclusions about low energy observables.

In the mass basis, we find the $Q_{quqd}^{(1)}$ term becomes 
\begin{equation}\label{eq:expandQquqdappendix}
\arraycolsep=1.4pt\def\arraystretch{1.8}
\begin{array}{rcl}
\mathcal{L}&=&-\frac{2}{3M_{\Z}^2}G_{ab}(G')^\dagger_{cd}\\
&&\times\left[(\bar u_R'u_L')_{ad}(\bar d_R d_L)_{bc}+(\bar u_R' d_L)_{ac}(\bar d_R u'_L)_{bd}\right]\, .
\end{array}
\end{equation}
A similar result is found for $Q_{quqd}^{(8)}$ with additional factors of $\lambda^A$.
Now we see the coefficient does not vanish for purely first generation couplings. We emphasize that these operators are not chirally invariant, in contrast to $Q_{quqd}^{(1,8)}$ when the latter are constructed solely from first generation quarks. As discussed in the main text, this feature follows from the generation non-diagonal property of the non-vanishing four-quark operators in this scenario. 

\bibliography{ref.bib}

\end{document}